# A Manifesto for Semantic Model Differencing

Shahar Maoz*, Jan Oliver Ringert**, and Bernhard Rumpe

Software Engineering
RWTH Aachen University, Germany
http://www.se-rwth.de/

**Abstract.** Models are heavily used in software engineering and together with their systems they evolve over time. Thus, managing their changes is an important challenge for system maintainability. Existing approaches to model differencing concentrate on heuristics matching between model elements and on finding and presenting differences at a concrete or abstract syntactic level. While showing some success, these approaches are inherently limited to comparing syntactic structures.

This paper is a manifesto for research on **semantic model differencing**. We present our vision to develop **semantic diff operators** for model comparisons: operators whose input consists of two models and whose output is a set of **diff witnesses**, instances of one model that are not instances of the other. In particular, if the models are syntactically different but there are no diff witnesses, the models are semantically equivalent. We demonstrate our vision using two concrete diff operators, for class diagrams and for activity diagrams. We motivate the use of semantic diff operators, briefly discuss the algorithms to compute them, list related challenges, and show their application and potential use as new fundamental building blocks for change management in model-driven engineering.

## 1 Introduction

Effective change management, a major challenge in software engineering in general and in model-driven engineering in particular, has attracted much research efforts in recent years (see, e.g., [5, 7, 12, 13, 15]). Due to iterative development methodologies, changing requirements, and bug fixes, models continuously evolve during the design, development, and maintenance phases of a system's lifecycle. Managing their changes using formal methods to follow their different versions over time is thus an important task. Fundamental building blocks for this task are *diff operators* one can use for model comparisons.

Existing approaches to model differencing concentrate on matching between model elements using different heuristics related to their names and structure and on finding and presenting differences at a concrete or abstract syntactic

* S. Maoz acknowledges support from a postdoctoral Minerva Fellowship, funded by the German Federal Ministry for Education and Research.
** J.O. Ringert is supported by the DFG GK/1298 AlgoSyn.

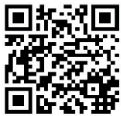

level. While showing some success, these approaches are also limited. Models that are syntactically very similar may induce very different semantics (in the sense of 'meaning' [9]), and vice versa, models that semantically describe the same system may have rather different syntactic representations. Thus, a list of syntactic differences, although accurate, correct, and complete, may not be able to reveal the real implications these differences have on the correctness and potential use of the models involved. In other words, such a list, although easy to follow, understand, and manipulate (e.g., for merging), may not be able to expose and represent the semantic differences between two versions of a model, in terms of the bugs that were fixed or the features (or new bugs...) that were added.

This paper is a manifesto for research on **semantic model differencing**. We present our vision to develop **semantic diff operators** for model comparisons: operators whose input consists of two models and whose output is a set of **diff witnesses**, instances of the first model that are not instances of the second. Such diff witnesses serve as concrete proofs for the real change between one version and another and its effect on the meaning of the models involved.

We demonstrate our ideas using two examples of concrete semantic diff operators, for class diagrams (CDs) and for activity diagrams (ADs), called *cddiff* and *addiff*, respectively. Given two CDs, *cddiff* outputs a set of diff witnesses, each of which is an object model that is an instance of the first CD and not an instance of the second. Given two ADs, *addiff* outputs a set of diff witnesses, each of which is a finite action trace that is possible in the first AD and is not possible in the second. Each operator considers the specific semantics of the relevant modeling languages, e.g., in terms of multiplicities, inheritance, etc. for CDs, and decision nodes, fork nodes, etc. for ADs.

In addition to finding concrete diff witnesses (if any exist), our operators can be used to compare two models and decide whether one model semantics includes the other model semantics (the latter is a refinement of the former), whether they are semantically equivalent, or whether they are semantically incomparable (each allows instances that are not allowed by the other). When applied to the version history of a certain model, such an analysis provides a semantic insight into the model's evolution, which is not available in existing syntactic approaches.

We have already implemented prototype versions of *cddiff* and *addiff*: all examples shown in this paper were computed by our prototype implementations. Section 4 gives a brief overview of the algorithms and tools we have used.

It is important not to confuse diffing with merging. Merging is a very important problem, dealing with reconciling the differences between two models that have evolved independently from a single source model, by different developers, and now need to be merged back into a single model (see, e.g., [2, 7, 12]). Diffing, however, is the problem of identifying the differences between two versions, for example, an old version and a new one, in order to better understand the course of a model evolution during some step of its development. Thus, diff witnesses are not conflicts that need to be reconciled. Rather, they are proofs of features that were added or bugs that have been fixed from one version to another along the history of the development process.

Finally, our vision of semantic diffing does not come to replace existing syntactic diffing approaches. Rather, it is aimed at augmenting and complementing existing approaches with capabilities that were not available before. As semantic differencing is so different from existing syntactic differencing approaches, it brings about new research challenges. We overview these challenges in Section 5.

The next section presents motivating examples, demonstrating the unique features of our vision. Section 3 presents a formal definition of a generic semantic diff operator and its specializations for CDs and ADs. Section 4 briefly describes the algorithms used to compute the two operators and their prototypes implementations, and Section 5 discusses new challenges emerging from our vision. Related work is discussed in Section 6 and Section 7 concludes.

## 2 Examples

We start off with a number of motivating examples, demonstrating the unique features of our vision.

**Example 1** Consider $cd1.v1$ of Fig. 1, describing a first version of a model for (part of) a company structure with employees, managers, and tasks. A design review with a domain expert has revealed two bugs in this model: first, employees should not be assigned more than two tasks, and second, managers are also employees, and they can handle tasks too.

Following this design review, the engineers created a new version $cd1.v2$, shown in the same figure. The two versions share the same set of named elements but they are not identical. Syntactically, the engineers added an inheritance relation between `Manager` and `Employee`, and set the multiplicity on the association between `Employee` and `Task` to 0..2. What are the semantic consequences of these differences?

Using the operator *cddiff* we can answer this question. $cddiff(cd1.v1, cd1.v2)$ outputs $om_2$, shown in Fig. 1, as a diff witness that is in the semantics of $cd1.v1$ and not in the semantics of $cd1.v2$; thus, it demonstrates that the bug of having more than two tasks per employee was fixed. In addition, $cddiff(cd1.v2, cd1.v1)$ outputs $om_1$, shown in Fig. 1, as a diff witness that is in the semantics of $cd1.v2$ and not in the semantics of $cd1.v1$. Thus, the engineers should perhaps check with the domain expert whether the model should indeed allow managers to manage themselves.

**Example 2** $cd5.v1$ of Fig. 2 is another class diagram from this model of company structure. In the process of model quality improvement, an engineer has suggested to refactor it by introducing an abstract class `Person`, replacing the association between `Employee` and `Address` by an association between `Person` and `Address`, and redefining `Employee` to be a subclass of `Person`. The resulting suggested CD is $cd5.v2$.

Using *cddiff* we are able to prove that despite the syntactic differences, the semantics of the new version is equivalent to the semantics of the old version, formally written $cddiff(cd5.v1, cd5.v2) = cddiff(cd5.v2, cd5.v1) = \emptyset$. The refactoring is indeed correct and the new suggested version can be committed.

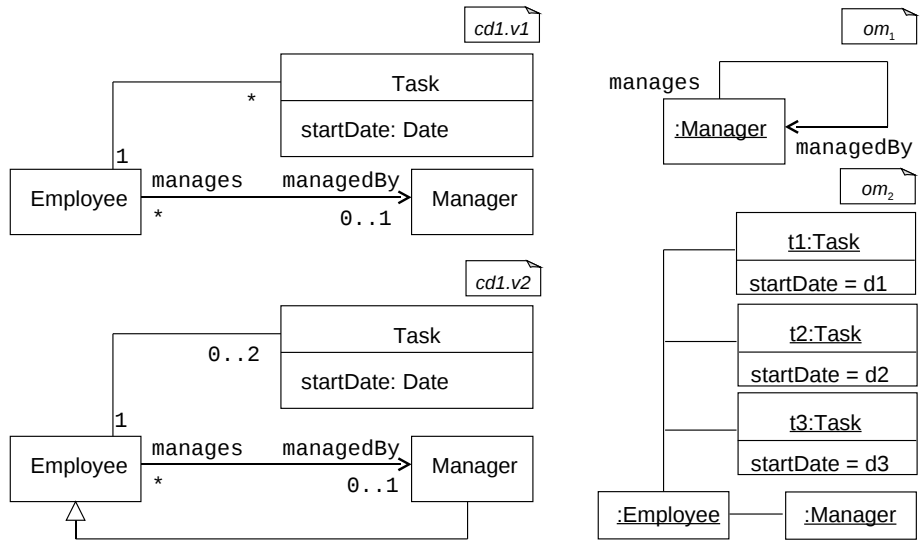

**Fig. 1.** $cd1.v1$ and its revised version $cd1.v2$, with example object models representing the semantic differences between them.

**Example 3** AD $ad.v1$ of Fig. 3 describes the company's workflow when hiring a new employee. Roughly, first the employee is registered. Then, if she is an internal employee, she gets a welcome package, she is assigned to a project and added to the company's computer system (in two parallel activities), she is interviewed and gets a manager report, and finally her payments are authorized. Otherwise, if the new employee is external, she is only assigned to a project before her payments are authorized.

After some time, the company deployed a new security system and every employee had to receive a key card. A revised workflow was created, as shown in $ad.v2$ of Fig. 3.

Later, a problem was found: sometimes employees are assigned to a project but cannot enter the building since they do not have a key card yet. This bug was fixed in the next version, $ad.v3$, shown in Fig. 4. Finally, the company has

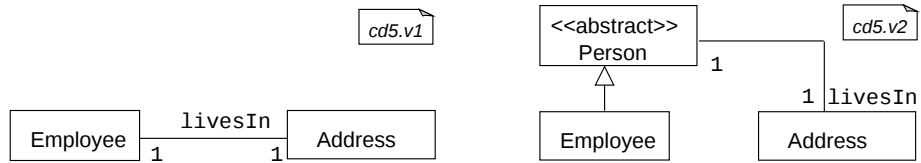

**Fig. 2.** Two example class diagrams of equivalent semantics

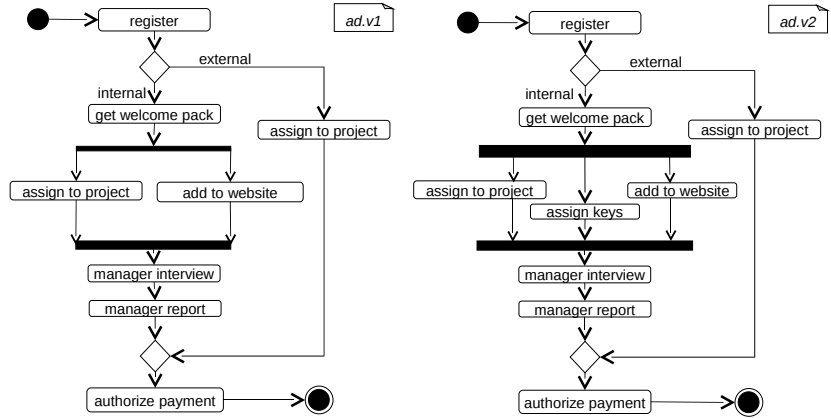

**Fig. 3.** Version 1 and version 2 of the hire employee workflow

decided that external employees should report to managers too. Thus, the merge between the two branches for internal and external new employees has moved 'up', in between the interview and the report nodes. The resulting 4th version of the workflow, $ad.v4$, is shown in Fig. 4

Comparing $ad.v1$ and $ad.v2$ using $addiff$ reveals that they are incomparable: some executions of $ad.v1$ are no longer possible in $ad.v2$, and some executions of $ad.v2$ were not possible in $ad.v1$. Moreover, it reveals that handling of internal employees has changed, but handling of external ones remained the same between the two versions.

Comparing $ad.v2$ and $ad.v3$ reveals that the latter is a refinement of the former: $ad.v3$ has removed some traces of $ad.v2$ and did not add new traces. In particular, $addiff(ad.v2, ad.v3)$ shows that the trace where a person is assigned to a project before she gets a security card was possible in $ad.v2$ and is no longer possible in $ad.v3$, i.e., it demonstrates that the bug was fixed.

Finally, comparing $ad.v3$ and $ad.v4$ using $addiff$ reveals that although hiring of external employees has changed between the two versions, hiring of internal employees did not: $addiff(ad.v3, ad.v4)$ contains a single trace, where the employee is external, not internal. That is despite the syntactic change of moving the merge node from after to before the report node, which is part of the handling of internal employees.

## 3 Formal Definitions

Consider a modeling language $ML = \langle Syn, Sem, sem \rangle$ where $Syn$ is the set of all syntactically correct (i.e., well-formed) expressions (models) according to some syntax definition, $Sem$ is a semantic domain, and $sem : Syn \to \mathcal{P}(Sem)$ is a function mapping each expression $e \in Syn$ to a set of elements from $Sem$ (see [9]).

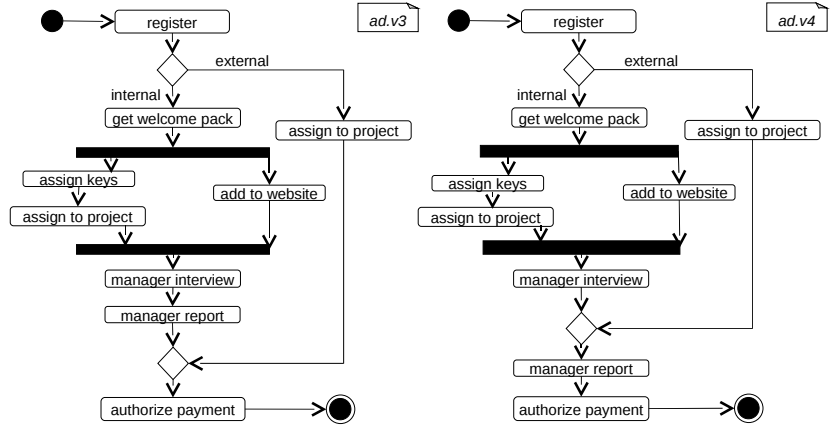

**Fig. 4.** Version 3 and version 4 of the hire employee workflow

The semantic diff operator $diff : Syn \times Syn \to \mathcal{P}(Sem)$ maps two syntactically correct expressions $e_1$ and $e_2$ to the (possibly infinite) set of all $s \in Sem$ that are in the semantics of $e_1$ and not in the semantics of $e_2$. Formally:

**Definition 1.** $diff(e_1, e_2) = \{s \in Sem | s \in sem(e_1) \land s \notin sem(e_2)\}$.

Note that $diff$ is not symmetric, $diff(e_1, e_1) = \emptyset$, and $diff(e_1, e_2) \cap diff(e_2, e_1) = \emptyset$. The elements in $diff(e_1, e_2)$ are called *diff witnesses*. We define specializations of $diff$ for CDs and ADs.

Our semantics of CDs is based on [8] and is given in terms of sets of objects and relationships between these objects. More formally, the semantics is defined using three parts: a precise definition of the syntactic domain, i.e., the syntax of the modeling language CD and its context conditions (we use MontiCore [11, 14] for this); a semantic domain - for us, a subset of the System Model (see [3, 4]) OM, consisting of all finite object models; and a mapping $sem : CD \to \mathcal{P}(OM)$, which relates each syntactically well-formed CD to a set of constructs in the semantic domain OM. For a thorough and formal account of the semantics see [4].

To make the operator *cddiff* computable and finite, we bound the number of objects in the witnesses we are looking for. Thus, we define a family of bounded operators. Formally:

**Definition 2 (cddiff).** $\forall k \geq 0, cddiff_k(cd_1, cd_2) = \{om | om \in sem(cd_1) \land om \notin sem(cd_2) \land |om| \leq k\}$, where $|om|$ is the maximal number of instances per class in om.

We use UML2 Activity Diagrams for the syntax of our ADs. In addition to action nodes, pseudo nodes (fork, decision, etc.), the language includes input and local variables (over finite domains), transition guards, and assignments. Roughly, the semantics of an AD is made of a set of finite action traces from an initial to a final node, considering interleaving execution of fork branches, the guards on decision nodes etc. (a formal and complete semantics of ADs is outside the scope of this paper).

In diffing ADs, we are looking only for shortest witnesses: diff traces that have another diff trace as prefix are not considered interesting. Formally:

**Definition 3 (addiff).** $addiff(ad_1, ad_2) = \{tr | tr \in sem(ad_1) \land tr \notin sem(ad_2) \land \nexists tr' : tr' \in sem(ad_1) \land tr' \notin sem(ad_2) \land tr' \sqsubseteq tr\}$.

## 4 Implementations and Applications

To evaluate our vision and demonstrate its feasibility, we have defined and implemented prototype versions of *cddiff* and *addiff*. Indeed, all examples shown in the previous section have been computed by our prototype implementations.

We compute a variant of *cddiff* using a transformation to Alloy [10]. Given two CDs, $cd_1$ and $cd_2$, we construct a single Alloy model consisting of the joint set of class signatures from the two CDs and a set of predicates that describe the relations between them in each of the CDs. We do not compute all instances of each CD and compare the two sets of instances; rather, we define a diff predicate, which specifies that all the $cd_1$ predicates hold and that at least one of the $cd_2$ predicates does not hold. We then use the Alloy Analyzer to compute instances of this diff predicate: these instances represent object models of the first CD that are not object models of the second CD. The transformation to Alloy considers the semantics of CDs, including multiplicities, inheritance, singleton and abstract classes etc.

Our implementation of *cddiff* can be used to compute diff witnesses, if any, or to show that no diff witnesses exist (up to a user-defined bound on the number of objects of each class in the model).

We compute a variant of *addiff* by modeling ADs as finite state machines, and defining a transformation to SMV [17]. Given two ADs, $ad_1$ and $ad_2$, we construct two SMV modules whose possible execution traces are exactly the set of possible traces of each of the ADs. We then use BDD-based algorithms, implemented using JTLV APIs [16], to find whether there are traces of $ad_1$ that are impossible in $ad_2$. The transformation to SMV and the algorithms used consider the semantics of ADs, including input variables, guarded branching in decision nodes, parallel interleaving execution following fork nodes, etc.

Our implementation of *addiff* can be used to compute diff witnesses, each of which is a finite trace which is a sequence of actions possible in one AD and not possible in the other (a trace includes the values of its input variables). If no such traces are found, we know that all traces of the first are also possible in the second, i.e., that the first is a refinement of the second. If, in addition, no such traces are found when reversing first and second, we know that the two ADs have equal semantics: their syntactic differences, if any, have no effect on their meaning.

We have integrated our implementations into Eclipse plug-ins. The plug-ins allow an engineer to compare two models from a project or two versions of a model from the history of a version repository. The engineer can then browse the diff witnesses that were found, if any.

Moreover, we have used *addiff* and *cddiff* to implement a `COMPARE` command, used to compare two selected models and output one of four answers: $\equiv$ if the two models are semantically equivalent, $<$ or $>$, if the second (first) is a semantic refinement of the first (second), and $<>$ if the two are incomparable, that is, if each of them allows instances (i.e., object models, traces) not possible in the other (in the case of *cddiff* the results of `COMPARE` are limited by the user-defined bound). `COMPARE` can be integrated with existing SVN history view, to provide a high-level semantic differencing summary of a model's evolution.

The details of the above transformations and algorithms for *cddiff* and *addiff*, and their related Eclipse plug-ins, are omitted from this workshop paper. We hope to present them in detail in follow-up publications.

## 5  Challenges

Semantic differencing is rather different from syntactic differencing approaches, so it raises a number of new research challenges.

### 5.1  Computation

Computing diff witnesses may not be algorithmically easy and sometimes even impossible. When computable, its complexity depends on the specific modeling language semantics at hand. For example, computing *cddiff* requires the use of a constraint solver (such as Alloy); to make it tractable, it must be bounded (see Section 3). Computing *addiff* requires a traversal of the state space induced by the ADs at hand. Depending on the use of fork nodes, input variables, and guards, this state space may be exponential in the size of the ADs themselves.

In general, depending on the available syntactic concepts and the semantics of the relevant modeling language, computing diff witnesses may be undecidable. In some cases, the set of computed witnesses may be sound but incomplete: all computed witnesses are indeed correct, but there may be infinitely many others that are harder to find. Thus, for each modeling language, a language specific diff operator needs to be defined and a new algorithm needs to be developed for its computation. Abstraction/refinement methodologies, partial-order reductions, and other approaches may be required in order to improve the efficiency of the computations and allow them to scale.

### 5.2  Presentation

To be useful, diff witnesses must be presented textually or visually to the engineer. Just like for computation, the presentation of diff witnesses is language specific; it depends on the specific modeling language of the models involved and its semantics. For example, for *cddiff*, differencing object models may be visually presented using generated object diagrams; for *addiff*, differencing traces may be visually presented on the ADs themselves, e.g., by coloring and numbering the nodes that participate in the diff trace on both diagrams, from the initial node up until the point where the two diagrams differ. Alternatively, one may use a collaboration diagram like notation, possibly with the aid of animation.

Moreover, as there may be (possibly infinitely) many diff witnesses, it is necessary to define sorting and filtering mechanisms, to select the 'most interesting' witnesses for presentation and efficiently iterate over them at the user's request.

### 5.3 Integration with syntactic differencing

Many works have suggested various syntactic approaches to model differencing (see Section 6). It may be useful to combine syntactic differencing with semantic differencing, for example:

– Extend the applicability of semantic diffing in comparing models whose elements have been renamed or moved in the course of evolution, by applying a syntactic matching before running a semantic diffing: this would result in a mapping plus a set of diff witnesses.
– Use information extracted from syntactic diffing as a means to localize and thus improve the performance of semantic diffing computations.

## 6 Related Work

The challenge of model change management and versioning has attracted much research efforts in recent years. In particular, many works have investigated various kinds of model comparisons. We review some of these briefly below.

[1] describes the difference between two models as a sequence of elementary transformations, such as element creation and deletion and link insertion and removal; when applied to the first model, the sequence of transformations yields the second. A somewhat similar approach is presented in [12] in the context of process models, focusing on identifying dependencies and conflicts between change operations. [7] presents the use of a model merging language to reconcile model differences. Comparison is done by identifying new/old MOF IDs and checking related attributes and references recursively. Results include a set of additions and deletions, highlighted in a Diff/Merge browser. [15] compares UML documents by traversing their abstract-syntax trees, detecting additions, deletions, and shifts of sub-trees.

As the above shows, some works go beyond the concrete textual or visual representation and have defined the comparison at the abstract-syntax level, detecting additions, removals, and shifts operations on model elements. However, to the best of our knowledge, no previous work considers model comparisons at the level of the semantic domain, as suggested in our vision.

Some works, e.g., [6, 18], use similarity-based matching before actual differencing. As our vision focuses on semantics, it assumes a matching is given. Semantic diffing can be applied after the application of matching algorithms.

## 7  Conclusion

In this paper we described our vision on semantic diff operators for model comparison, as new fundamental building blocks for change management in model-driven engineering. We motivated our vision with examples, and gave a brief overview of the formal background and the algorithms used in our prototype implementations. Finally, we listed new research challenges that emerge from our vision, related to the computation and presentation of semantic model differences.

**Acknowledgments** We thank Assaf Marron and the anonymous reviewers for their comments on a draft of this paper.